\documentclass[11pt,titlepage]{article}
\usepackage{epsf}
\usepackage{latexsym}
\newcommand{\Square}{
\hspace{-4mm}
\setlength{\unitlength}{1.0mm}
\begin{picture}(2,3)
\linethickness{2.0mm}
\put(0,1){\line(1,0){2}}
\end{picture}}
\advance\textheight by \topskip
\newlength{\bredde}
\renewcommand{\to}{\rightarrow}
\newcommand{\beq}[1]{\begin{equation}\label{#1}}
\newcommand{\eeq}{\end{equation}}
\newcommand{\bea}[1]{\begin{eqnarray}\label{#1}}
\newcommand{\eea}{\end{eqnarray}}
\newcommand{\rf}[1]{(\ref{#1})}
\newcommand{\vev}[1]{ {\langle #1 \rangle} }
\def\slash#1{\settowidth{\bredde}{$#1$}\ifmmode\,\raisebox{.15ex}{/}
\hspace*{-\bredde}#1\else$\,\raisebox{.15ex}{/}\hspace*{-\bredde} #1$\fi}
\def\e{{\rm e}}
\def\ln{{\rm ln}}
\begin{document}
\thispagestyle{empty}
\begin{titlepage}
\addtolength{\baselineskip}{.7mm}
\thispagestyle{empty}
\begin{flushright}
NBI-HE-97-21\\
NORDITA-97/23\\
hep-lat/9705004\\
\end{flushright}
\vspace{10mm}
\begin{center}
{\Large{\bf
Singularities of the Partition Function for the \\
Ising Model Coupled to 2d Quantum Gravity
}}\\[15mm]
{\sc 
J.~Ambj{\o}rn$^{\dagger}$,
K.~N.~Anagnostopoulos$^{\dagger}$ and
U.~Magnea$^{\ddagger}$ 
} \\
\vspace{10mm}
{\it
${}^{\dagger}$The Niels Bohr Institute \\
${}^{\ddagger}$NORDITA\\
Blegdamsvej 17, DK-2100 Copenhagen \O, Denmark}\\[6mm]
\vspace{13mm}
{\bf Abstract}\\[5mm]
\end{center}

We study the zeros in the complex plane of the partition function for
the Ising model coupled to $2d$ quantum gravity for complex magnetic
field and real temperature, and for complex temperature and real
magnetic field, respectively. We compute the zeros by using the exact
solution coming from a two matrix model and by Monte Carlo simulations
of Ising spins on dynamical triangulations. We present evidence that
the zeros form simple one-dimensional curves in the complex plane, and
that the critical behaviour of the system is governed by the scaling
of the distribution of the singularities near the critical
point. Despite the small size of the systems studied, we can obtain a
reasonable estimate of the (known) critical exponents.

\noindent
\vfill
\renewcommand{\thefootnote}{}
\footnotetext{
e-mail addresses: ${}^\dagger$ ambjorn, konstant, 
${}^\ddagger$ umagnea@nbi.dk}
\end{titlepage}
\newpage
\renewcommand{\theequation}{\arabic{section}.\arabic{equation}}
\renewcommand{\thefootnote}{\fnsymbol{footnote}}
\setcounter{footnote}{0}
\section{Introduction}

The understanding of the phase transition of a statistical mechanical
model has long been connected to the study of its partition function
zeros. This line of research was pioneered by Yang and Lee \cite{YL},
and has subsequently been pursued by many authors (for recent work see
\cite{S} and references therein).  In the thermodynamic limit, the
zeros of the partition function of such a system in the complex
coupling space accumulate infinitely close to the critical coupling on
the real axis.  This defines, in the infinite volume limit,
disconnected regions in the complex plane, with different analytic
structure for the thermodynamic functions which describe the phases of
the system.

The critical behaviour in the neighbourhood of a continuous phase
transition can be extracted from the behaviour of the density of zeros
near its singular points \cite{F,ABE,OKSK}. In the best-understood
models, these points occur at the ends of lines on which the zeros
accumulate in the infinite volume limit. Such singular points, even
when they occur at non--physical values of the coupling, can be
considered as ordinary critical points with distinctive critical
exponents \cite{FYL} (see also \cite{DH}). For a physical phase
transition, these singularities coincide with the (real) critical
coupling of the system.

Substantial progress, also from a practical point of view, has been
made by applying renormalization group arguments to the motion of zeros,
thereby deriving finite size scaling relations by which the
critical exponents can be computed \cite{IPZ}. This finite-size scaling
technique applied to complex zeros has been proven powerful in
numerical computations of critical exponents.

A special class of statistical systems are spin systems defined on
random surfaces. In addition to being interesting systems {\it per
se}, they have received special attention for their relation to
conformal matter systems coupled to two dimensional quantum
gravity. At criticality, the spin systems are described by conformal
field theories (CFT) coupled to gravity. The critical exponents of
such conformal theories have been found by Knizhnik, Polyakov and
Zamolodchikov (KPZ) to be simply related to the critical exponents of
the same CFT defined in flat space \cite{KPZ}. In particular, the
critical Ising model coupled to $2d$ gravity can be recast in the form
of a $c=1/2$ conformal field theory coupled to $2d$ gravity. In
addition, the equivalence of the Ising model on a random surface to a
hermitean two matrix model, discovered and studied by Kazakov and
Boulatov \cite{K,BK}, renders it exactly solvable.  The critical
exponents were first computed this way, outside of the CFT
context. The system is found to have a third order phase transition
from a high temperature disordered phase to a low temperature ordered
phase. The coupling between matter and the geometry of the surface is
found to change the critical exponents associated to geometry only at
the phase transition.

The location of the partition function zeros of the Ising model in the
complex fugacity plane, the so called Lee--Yang zeros, in the presence
of gravity has been studied by Staudacher \cite{Staudacher}.  The
somewhat surprising discovery that the Lee--Yang zeros lie on the unit
circle in the complex fugacity plane {\it for each lattice size $N$}
will be confirmed below for slightly bigger lattices than those
considered in \cite{Staudacher}.  As discussed by Staudacher, the
Lee--Yang theorem \cite{YL} guarantees that the roots of the partition
function of the Ising model on a fixed lattice $Z_{{\rm flat}}(G_N)$
for a large class of not too pathological lattices $G_N$ will lie on
the unit circle, but it is not a priori expected that the roots of the
partition function $Z_N$, obtained by summing the $Z_{{\rm
flat}}(G_N)$'s over all dynamical random lattices ${G_N}$,
\begin{equation}
Z_N = \sum_{\{G_N\}} Z_{{\rm flat}}(G_N), \label{eq:summ}
\end{equation}
will also be located on the unit circle. The summation in
(\ref{eq:summ}) corresponds to an integration over the additional
quantum degree of freedom introduced by a fluctuating metric.  The
Lee--Yang theorem for this case remains unproven.  {\it A priori} the
zeros can be located on different curves or even on sets with more
complicated topology, or on two-dimensional regions. The latter cases
have been observed for the complex temperature zeros of $2d$ Ising and
Potts models on hierarchical lattices \cite{DDI}, for anisotropic
lattices \cite{SKSC}, and interestingly, also for a regular lattice made 
of different polygons \cite{rs2}. (See also ref.~\cite{DamL}, where a
very interesting, albeit somewhat inconclusive attempt at studying the
partition function zeros of Ising spin glasses is made, and ref.~\cite{aper},
where the zeros for aperiodic systems were studied.)

In this paper we use the same method as in \cite{Staudacher} in order
to compute the exact partition function for systems with up to 14
squares. We compute the Lee--Yang zeros of the partition function and
confirm that they are located on the unit circle of the complex
fugacity plane. We also compute the zeros of the partition function
for complex temperature (Fisher zeros) and find that they move on
curves in the complex temperature plane as the lattice size is
varied. This indicates that in the infinite volume limit, the zeros
accumulate to form dense sets on these curves.  We test the scaling
relations derived by Itzykson, Pearson and Zuber in the context of
regular lattices \cite{IPZ} and extract from there some (combinations
of) critical exponents.  These scaling relations are based on the
assumption that the spin--spin correlation length diverges at the
critical point.  Despite the small size of the systems studied, we
find reasonable agreement with the critical exponents given in
\cite{K,BK}. This result is expected from the non--trivial fact that
the matter correlation length diverges at the phase transition
\cite{corr} (see \cite{ha} for an example where this is not true).
This way we provide strong evidence that the phase transition is
governed by the singularities of the distribution densities at $H=0$
and $\beta_c$ of the Lee--Yang and Fisher zeros respectively as is the
case for the Ising model on a fixed lattice. The pattern of zeros we
observe confirms the observation in \cite{BJ} that no
antiferromagnetic transition occurs for the Ising model on a square
lattice when the spins are placed on the faces of the lattice. Using
duality, we can study the pattern of the Fisher zeros of the dual
model where the spins are placed on the vertices.  We observe that the
set of Fisher zeros maps onto itself under ${\tilde \beta}\to -{\tilde
\beta}$.  This indicates the existence of an antiferromagnetic
transition whose critical exponents are identical to the ones of the
ferromagnetic transition\footnote{We thank Desmond Johnston for
pointing this out to us.}. This confirms the results reported in
\cite{BJ}.

One of the methods for computing complex zeros is provided by Monte
Carlo simulations and histogramming techniques. These techniques have
been applied successfully to systems defined in flat space
\cite{MCZ,MAR}.  They have also been tried on several systems coupled
to gravity (random surfaces coupled to extrinsic curvature, Potts
models, 4d quantum gravity \cite{MBA}), but it proved
difficult to observe the partition function complex zeros. Here, we
apply the method to Ising spins on dynamical triangulations \cite{bj},
and we are able to observe the zeros lying closest to the real axis
for a range of lattice sizes with up to 256 vertices. In general, it
was easier to find the Lee--Yang zeros than the Fisher zeros. In the
latter case, we were able to locate only the first zero for any given
lattice size. After determining the position of the zeros as a
function of lattice size, we test their scaling and the critical
exponents computed are in reasonable agreement with their exactly
known values.

In section~\ref{s:m} we define our model and review the expected
scaling behaviour of the zeros on the assumption of a diverging matter
correlation length. In section~\ref{s:z} we describe in detail the
methods we use in order to compute the zeros, and in section~\ref{s:v} we
test their finite size scaling behaviour. Section~\ref{s:c}
contains some concluding remarks.

\section{The model}
\label{s:m}

We will study one Ising model on a dynamical square or triangular
lattice. In the former case the spins are located on the $N$ faces
(squares) of the lattice, in the latter on the $N_v$ vertices.
The partition function for a fixed lattice $G_N$ is given by
\begin{equation}
\label{*1}
Z_{{\rm flat}}(G_N,\beta,H) = 
\sum_{\{\sigma\}} \e^{{\beta}\sum_{\vev{i,j}}\sigma_i
\sigma_j + H \sum_i \sigma_i}\, , \label{eq:ZGN}
\end{equation}
and the partition function $Z(N,\beta,H)$ for the model coupled to
quantum gravity is obtained by summing $Z_{{\rm flat}}(G_N,\beta,H)$ over
all lattices $G_N$ with spherical topology and $N$ faces
\begin{equation}
\label{*2}
Z(N,\beta,H) = \sum_{\{G_N\}} Z_{{\rm flat}}(G_N,\beta,H)\, .
\end{equation}  
In eq.~(\ref{eq:ZGN}) $\sum_{\vev{i,j}}$ denotes a sum over the 
neighbouring pairs of faces (in case of a square lattice) or vertices 
(in case of a triangular lattice) of $G_N$.
$\sum_{\{\sigma\}}$ denotes the sum over all the possible spin
configurations, $\beta$ is the inverse temperature and $H$ the
magnetic field. $\sum_{\{G_N\}}$ is the discrete analogue of summing
over all possible metrics up to diffeomorphisms on a spherical surface
of fixed volume, since $G_N$ defines a metric on the discretized
surface if we define each link to have the same fixed length.  The
class of lattices we consider include degenerate ones which have
double links and vertices of order one.

The system \rf{*2} undergoes a third order phase transition whose
universal properties are independent of the microscopic details of the
lattice \cite{K,BK}. For the square lattice, the critical inverse
temperature is given by $\beta_c = \ln 2 \approx 0.69314718$, and for
the triangular lattice, $\beta_c = \frac{1}{2}\ln
\frac{13+\sqrt{7}}{14-\sqrt{7}}\approx 0.16030370$ \cite{BJ}. The
critical exponents are given by \cite{BK}
\begin{eqnarray}
\beta   &=& {1/2}\\
\gamma  &=& 2\\
\delta  &=& 5\\
\nu d_H &=& 3 \, ,
\end{eqnarray}
which are defined in the usual way by the behaviour 
of the magnetization $M\sim |\beta-\beta_c|^\beta$, 
$M\sim H^{1/\delta}$ and  
the magnetic susceptibility $\chi_M \sim|\beta-\beta_c|^{-\gamma}$ in
the critical region. $\nu$ is the spin--spin
correlation length exponent $\xi \sim |\beta-\beta_c|^{-\nu}$, 
and $d_H$ denotes the Hausdorff dimension of space.

For a given lattice $G_N$, the partition function \rf{*1} can be
written (up to a multiplicative constant) as a polynomial
\beq{*3}
Z_{{\rm flat}}(G_N,\beta,H) = \sum_{m=0}^{N}\sum_{n=0}^{N_l} C_{mn} y^m c^n\, ,
\eeq
where $c= \e^{-2\beta}$, $y=\e^{-2H}$, $N$ is the number of faces in
$G_N$ (when we put the spins on the faces) and $N_l$ the number of
links. $y$ is called the fugacity. In the following we will also use
the notation $u = \e^{-4\beta}$ and $K=u-u_c$.

Eq. \rf{*3} is a polynomial in two variables, and its analytic
structure is entirely determined by its zeros in the complex plane.
If we fix the temperature $\beta$, $Z_{{\rm flat}}(G_N,\beta,H)$ is a
polynomial in $y$ and its zeros were shown by Yang and Lee \cite{YL}
to lie on the unit circle in the complex fugacity plane. We expect the
theorem to be true for a large class of (not too pathological) lattices
$G_N$. Then one can write the free energy of the system as
\beq{*4a} F_{{\rm flat}}(G_N,\beta,y) = - \ln
\prod_{k=0}^{N}(y-y_k(\beta))\, , 
\eeq 
where a factor $\beta$ has been absorbed into $F_{{\rm
flat}}(G_N,\beta,y)$ and an additive constant has been suppressed
($y\equiv\e^{-2H}$). Here $y_k(\beta)$ are the zeros of $Z_{{\rm
flat}}(G_N,\beta,H)$ in the $y$ plane, called the Lee--Yang zeros.  In
the thermodynamic limit these zeros form dense sets on lines, which
are Stokes lines for $F_{{\rm flat}}(G_{\infty },\beta,y)$. All the
relevant information is then encoded in the density of zeros
$\rho_{YL}(\beta,\theta)$ ($H=i\theta$). 
The free energy per spin in the limit of
infinite volume $N \to \infty$ is then given by
\beq{*5}
F_{{\rm flat}}(G_{\infty },\beta,y) = -\int_{-\pi}^\pi d\theta\, 
\rho_{YL}(\beta,\theta) \, 
\ln(y-\e^{-2i\theta})\, .
\eeq
Here $-2\theta$ denotes the polar angle on
the unit circle of the partition function zero in the complex
$y$-plane. This notation is motivated by the Lee--Yang theorem. The
singularities of the partition function are expected to occur at the
ends of the lines on which the partition function zeros condense. In
the case of \rf{*1}, in the high temperature phase the zeros form a
gap $\theta_0>0$ such that $\rho_{YL}(\beta,\theta) = 0$ for
$|\theta|<\theta_0$.  The points $y_0=\e^{\pm 2i\theta_0}$ are the
Lee--Yang edge singularities and they can be regarded as  conventional
critical points \cite{FYL}. The (only) characteristic critical exponent
$\sigma$ is
\beq{*6}
\rho_{YL}(\beta,\theta)\sim (\theta-\theta_0)^\sigma\, ,
\eeq
which implies that $M\sim (\theta-\theta_0)^\sigma$. In \cite{Cardy}
it was shown that $\sigma=-1/6$ in two dimensions.

As we approach the critical temperature from above, the gap closes
($\theta_0 \to 0$) and the zeros pinch the real axis at $y=1$. This
signals the onset of the phase transition, since the
partition function has different analytic behaviour in the two
disconnected parts of the $y$-plane. The information on the universal
scaling of the partition function is given by the scaling of
$\rho_{YL}(\beta,\theta)$ near $\theta=0$. At $\beta=\beta_c$,
$\theta_0=0$ and if we define $M \sim H^{1/\delta}$ we obtain
\beq{*6a}
\rho_{YL}(\beta_c,\theta) \sim \theta^{1/\delta}\, .
\eeq
Other critical exponents can be derived from scaling relations
discussed in \cite{IPZ} and later in this section.  For a finite
system the zeros are never on the positive real axis since $C_{mn}$ is
positive and $c>0$ for real $\beta$ in eq.~\rf{*3}.  When the system
is critical, i.e. when the correlation length $\xi\sim L$ ($L$ is the
linear size of the system), the zeros approach the real axis infinitely 
close as $L\to\infty$ and one can apply finite size scaling in
order to extract critical exponents.

If we fix $y$ in eq.~\rf{*3}, we obtain a polynomial of order $N_l$ in
$c$. Its zeros in the $c$-plane are the so-called Fisher zeros
\cite{F}. For many systems the Fisher zeros are located on curves
${\cal C}$, but this is not necessarily true in general.  For the $2d$
(zero-field) Ising model on a square lattice, these curves ${\cal C}$
are two intersecting circles (both in the $c$-plane and in the ${\rm
tanh}(\beta)$ plane).  One of the curves intersects the real positive
axis at the physical critical point. Similarly to the case of
Lee--Yang zeros, the Fisher zeros condense on these two curves in the
thermodynamic limit and the free energy per spin is given, in this
limit, by their density $\rho_F(\beta,H)$:
\beq{*7}
F_{{\rm flat}}(G_{\infty},\beta,H) = - 
\int_{\cal C}\, d\beta'  \rho_F(\beta',H)
\,\ln(c(\beta)-\e^{-2\beta'})\, , 
\eeq
where $\beta'$ is a complex number. Near the critical temperature and at $H=0$,
\beq{*7a}
\rho_F(\beta, H=0)\sim |\beta-\beta_c|^{-\alpha + 1}\, ,
\eeq
which implies that $F_{{\rm flat}}(G_{\infty },\beta,H=0) \sim
|\beta-\beta_c|^{-\alpha + 2}$. Therefore $\alpha$ is the specific
heat exponent.

The possibility of extracting scaling exponents from the study of
complex zeros of the partition function relies on their scaling
behaviour under the renormalization group \cite{IPZ}.  Applying finite
size scaling is a convenient method often used in numerical
simulations. Simple scaling arguments \cite{IPZ} give the position of
the $j$th Lee--Yang zero as
\beq{*8}
H_j^2 N^{2\beta\delta/(\nu d_H)} = f_j(K N^{1/(\nu d_H)})\, .
\eeq
where we have substituted $N^{1/d_H}$ for the linear size $L$ of the
system ($N$ is the volume and $d_H$ is the Hausdorff
dimension). $f_j$ is an analytic
function and the Lee--Yang theorem implies that $f_j(0)<0$. 
If we invert the above relation
we obtain the positions of the Fisher zeros. Hence we can deduce that
the $j$th  Lee--Yang zero will scale as
\beq{*9}
H_j \sim N^{-\beta\delta/(\nu d_H)}\, ,
\eeq
and the $j$th Fisher zero as 
\beq{*10}
K_j \sim N^{-1/(\nu d_H)}\, .
\eeq
The gap of the Lee--Yang edge singularity will scale as
\beq{*10a}
H_0^2 \sim  -C K^{2\beta\delta} \, ,
\eeq
where $C>0$ and $H_0\equiv i\theta_0$. Conversely, in the scaling region
the dependence of the $j$th Fisher zero on the value 
of the (real) magnetic field will be 
\beq{*11}
K_j \sim \e^{\frac{i\pi}{2\beta\delta}}
\left( {{H}\over{\sqrt{C}}} \right)^{1/(\beta\delta)}\, .
\eeq
Eq.~\rf{*11} implies that the trajectories of the motion of the Fisher
zeros in the $K$ plane with varying real nonzero $H$ will form an angle 
\beq{*12}
\psi = \frac{\pi}{2\beta\delta} 
\eeq
with the real $K$ axis. Using eq.~\rf{*8} a  
stronger scaling relation can be derived:
\beq{*13}
H_j^2(N/j)^{2\delta/(\delta+1)} =  F\left( K (N/j)^{1/(\nu d_H)}\right) \, ,
\eeq
where $F$ is a universal analytic scaling function such that $F(0)<0$, $H_j$
is the $j$th Lee--Yang zero. Similarly
\beq{*14}
K_j=\left( {j\over N} \right)^{1/(\nu d_H)} F^{-1}(0)\, ,
\eeq
where $K_j$ is the $j$th Fisher zero. 

In \cite{corr}, we provided evidence that there exists a divergent
correlation length associated to matter fields coupled to gravity in
the range of central charge $0<c<1$.  Based on these results, we
expected that the observed zeros, for large enough lattices, might
show a behaviour compatible with the scaling hypothesis. It is the
main purpose of this work to provide evidence that this is indeed the
case.
 
\section{Computation of Complex Zeros}
\label{s:z}

Our computation of Lee--Yang zeros (real $\beta$, complex $H$) and
Fisher zeros (real $H$, complex $\beta$) of the partition function
$Z(N,\beta,H)$ was done using exact results from matrix models and
Monte Carlo multihistogramming techniques. We describe these methods
in this section.

\subsection{Exact determination of partition functions} 

The exact solution of the Ising model on a square dynamical lattice 
came from solving the planar limit of the two matrix model defined in 
\cite{K,BK}. It was noticed that in this limit the free energy of the model 
\bea{*15}
F(n,g,\beta,H) &=&
\ln \Bigg( \int d^{n^2} \phi_1 d^{n^2} \phi_2\nonumber\\
&&\quad\exp\left[-{\rm Tr}\left( \phi_1^2 + \phi_2^2 - 2c \phi_1^2\phi_2^2
+(g\e^H/n)\phi_1^4 + (g\e^{-H}/n)\phi_2^4 \right)\right] \Bigg)
\eea
where $\phi_{1,2}$ are $n\times n$ hermitean matrices, 
equals the grand canonical partition function $Z(g,\beta,H)$ 
for the Ising model coupled to gravity:
\beq{*16}
\lim_{n\to\infty} \frac{1}{n^2} F(n,g,\beta,H) = Z(g,\beta,H)\equiv
 \sum_{N=1}^\infty \tilde{c}^N Z(N,\beta,H)\, ,
\eeq
where $\tilde{c}=-4gc/(1-c^2)^2$. The solution for $Z(g,\beta,H)$ is
given by \cite{K,BK}
\beq{*17}
Z(g,\beta,H) = \frac{1}{2}\ln\left[\frac{z(g)}{g}\right] +
\frac{1}{2g^2}\int_0^{z(g)}\frac{dz'}{z'} g(z')^2 -
\frac{1}{g}\int_0^{z(g)}\frac{dz'}{z'} g(z')\, ,
\eeq
where
\beq{*18}
g(z)=\frac{1}{9}c^2 z^3 + \frac{1}{3} z 
\left[ \frac{1}{(1-z)^2}-c^2+\frac{2z(\cosh H - 1)}{(1-z^2)^2} 
\right]\, .
\eeq
Using eq.~\rf{*17} and eq.~\rf{*18} we can expand $Z(g,\beta,H)$ in
powers of $g$ and read off the coefficients $Z(N,\beta,H)$ (corresponding 
to lattice size with $N$ squares) in eq.~\rf{*16}. 

We have done this using the symbolic manipulation program {\it
Mathematica} for $N\leq 14$. We obtain 
\beq{*18a}
Z(N,\beta,H)=
c^{-N}y^{-\frac{N}{2}}\sum_{m=0}^{N}\sum_{n=0}^{4\left[\frac{N}{2}\right]}
D_{mn}\, y^m\, c^n\, ,
\eeq
where $\left[ . \right]$ denotes integer part.  The same calculation
for $N\leq 6$ was done in \cite{Staudacher} and our results are in
complete agreement. Then the roots of $Z(N,\beta,H)$ for either fixed
$\beta$ or fixed $H$ were computed numerically. Note that, apart from
a trivial multiplicative factor, the partition function is a
polynomial and has a finite number of zeros in the complex $c$ and $y$
planes.

\begin{figure}[b]
\centerline{\epsfxsize=4.0in \epsfysize=2.67in \epsfbox{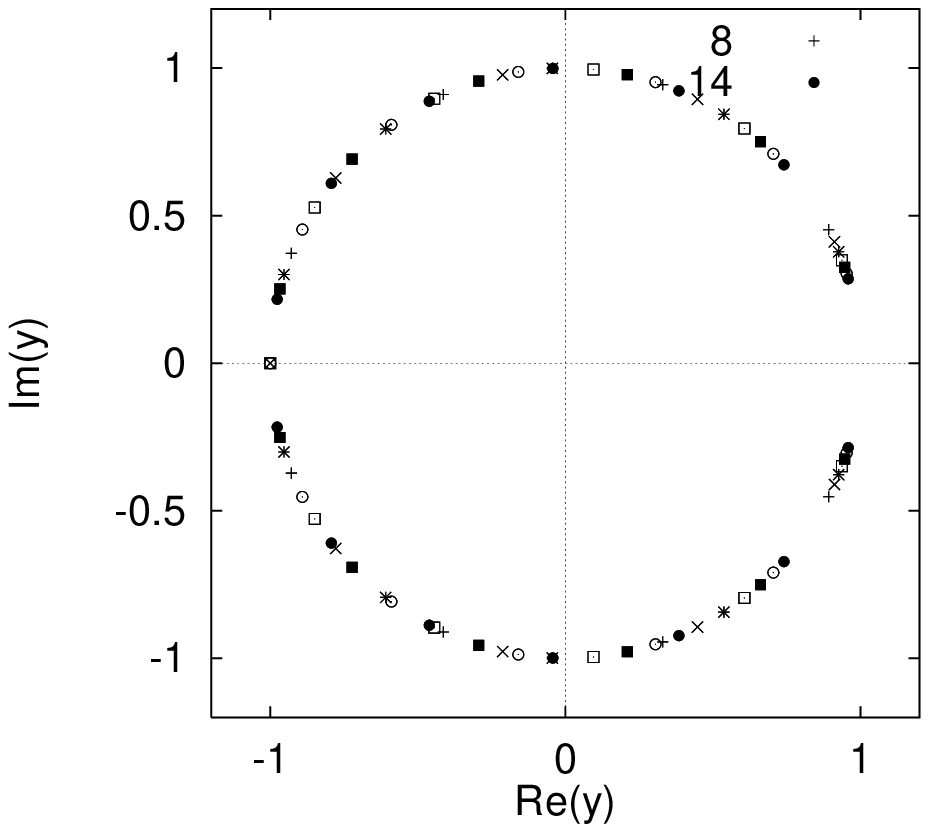}}
\caption{Lee--Yang zeros in the complex fugacity ($y=\e^{-2H}$) plane
for the Ising model on square dynamical lattices of varying size $N=$
8($+$), 9($\times$), 10($\ast$), 11($\Box$),
12(\protect\Square), 13($\circ$), 14($\bullet$) at the
bulk critical temperature.}
\label{f:1}
\end{figure}

As already discussed above, and in \cite{Staudacher} for $N\leq 6$,
the Lee--Yang zeros can be seen, somewhat surprisingly, to lie on the
unit circle in the complex $y$-plane. No Lee--Yang theorem has so far
been proven for the partition function of the dynamical lattice.
Fig.~\ref{f:1} shows the results for lattice sizes $8\le N\le 14$, at
the (bulk) critical temperature.  This figure clearly shows that the
zero closest to the real positive axis approaches the point $y=1$ as
we increase the lattice size, indicating a vanishing gap at $\beta_c$
as $N \to \infty$.
We will study this approach quantitatively using finite size scaling
in the next section.

\begin{figure}[hb]
\centerline{\epsfxsize=4.0in \epsfysize=2.67in \epsfbox{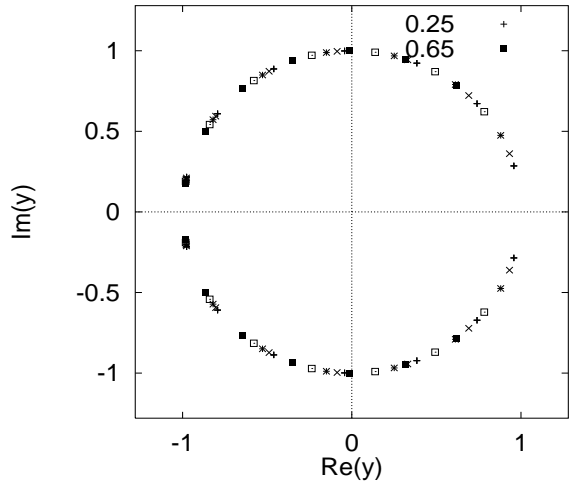}}
\caption{Lee--Yang zeros in the complex fugacity plane for a square lattice 
with $N=14$ at the critical point ($c_c=\e^{-2\beta_c}=$ 0.25($+$)) and in 
the hot phase ($c=$ 0.35($\times$), 0.45($\ast$), 0.55($\Box$),
0.65(\protect\Square)).} 
\label{f:2}
\end{figure}

In Fig.~\ref{f:2} we show the motion of Lee--Yang zeros with varying
$\beta$ for the $N=14$ lattice as we approach $\beta_c$ from the hot
phase. The zeros close up towards the point $y=1$ ($H=0$) on the real
axis, reflecting the expected vanishing of the gap in
$\rho_{YL}(\beta,\theta)$ for $\beta \to \beta_c$ in the infinite volume
limit.  

\begin{figure}[htb]
\centerline{\epsfxsize=4.0in \epsfysize=2.67in \epsfbox{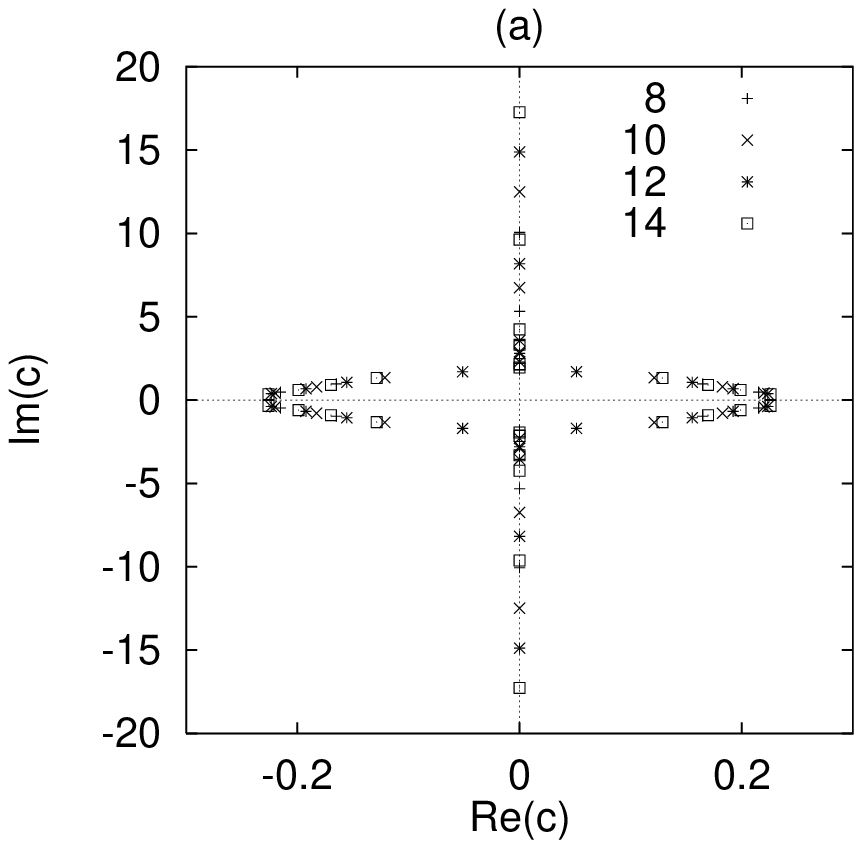}}
\centerline{\epsfxsize=4.0in \epsfysize=2.67in \epsfbox{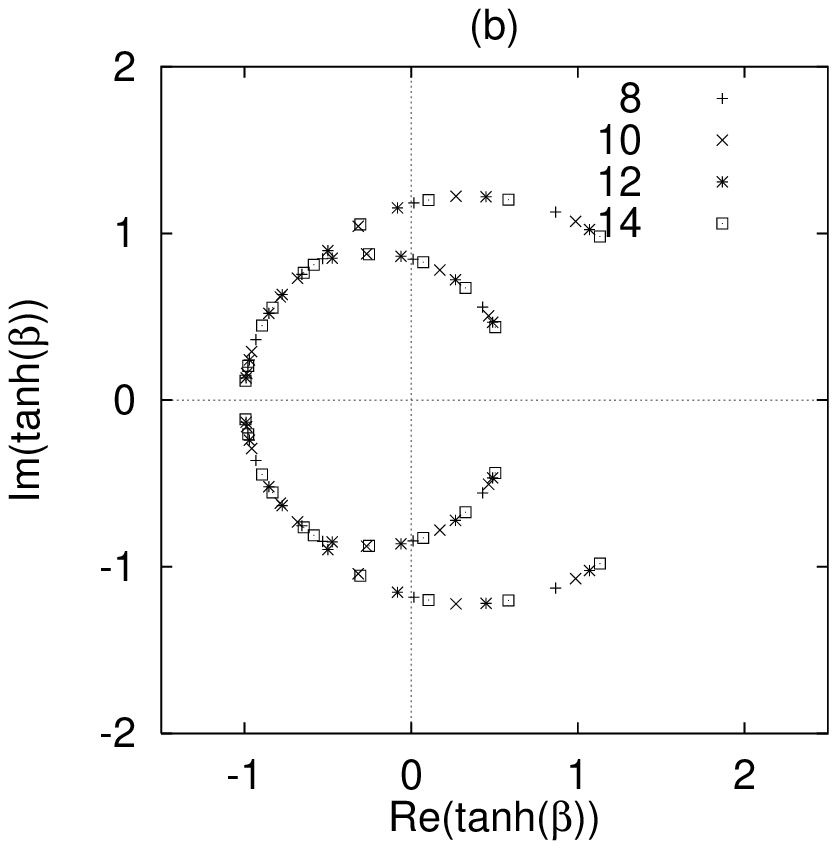}}
\caption{(a) Fisher zeros in the complex $c$ plane ($c=\e^{-2\beta}$)
for the Ising model with spins on the faces of square dynamical lattices 
of varying size
($N=8,\ 10,\ 12,\ 14$) at $H=0$.  The zeros move on arcs and on the
imaginary axis. The zeros flow towards $\pm i\infty$ on the ${\rm
Im}(c)$ axis, and towards the critical points $c=(1/4,0)$ and
$c=(-1/4,0)$ on the real axis as the thermodynamic limit is
approached.  The physical critical point is $c=1/4$. (b) The
trajectories in (a) shown in the complex ${\rm tanh}(\beta)$
plane. Recalling the duality relation ${\tilde c}={\rm tanh}(\beta)$,
this corresponds to Fig.~3(a) for the model with spins on the vertices
(with ${\tilde c}$ related to the dual inverse temperature
$\tilde{\beta}$ by ${\tilde c}=\e^{-2\tilde{\beta}}$).  The zeros
approach the points $(-1,0)$, $(3/5,0)$ and $(5/3,0)$ as $N \to
\infty$. The point ${\tilde c}=3/5$ corresponds to the ferromagnetic
and the point ${\tilde c}=5/3$ to the antiferromagnetic transition of
the model. The zeros are mapped onto each other under ${\tilde c}\to
1/{\tilde c}.$}
\label{f:3}
\end{figure}

\begin{figure}[htb]
\centerline{\epsfxsize=3.0in \epsfysize=2.0in \epsfbox{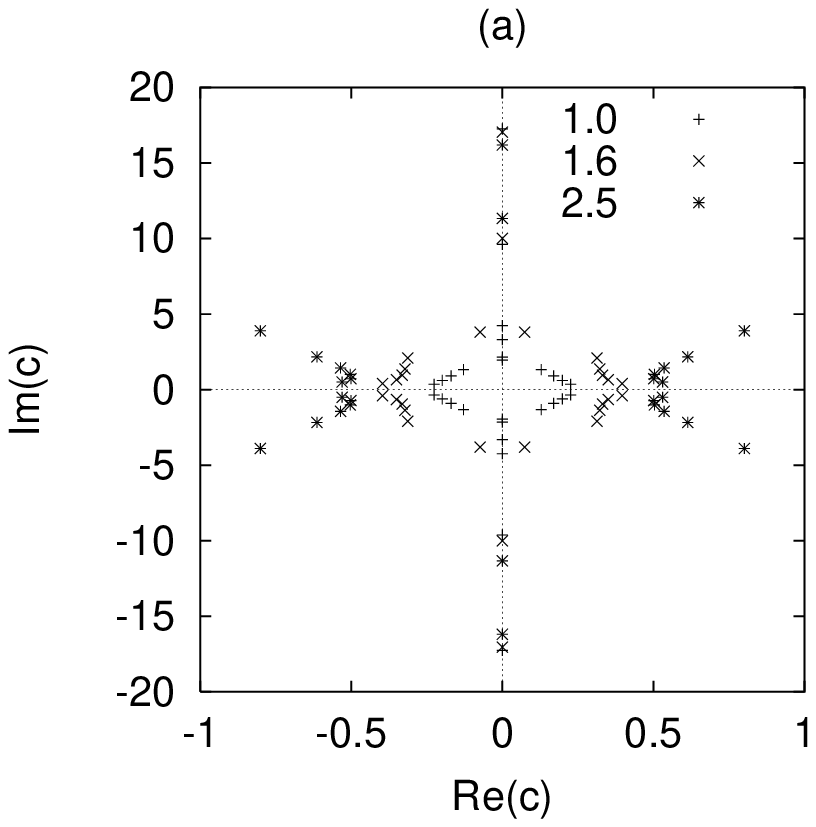}}
\centerline{\epsfxsize=3.0in \epsfysize=2.0in \epsfbox{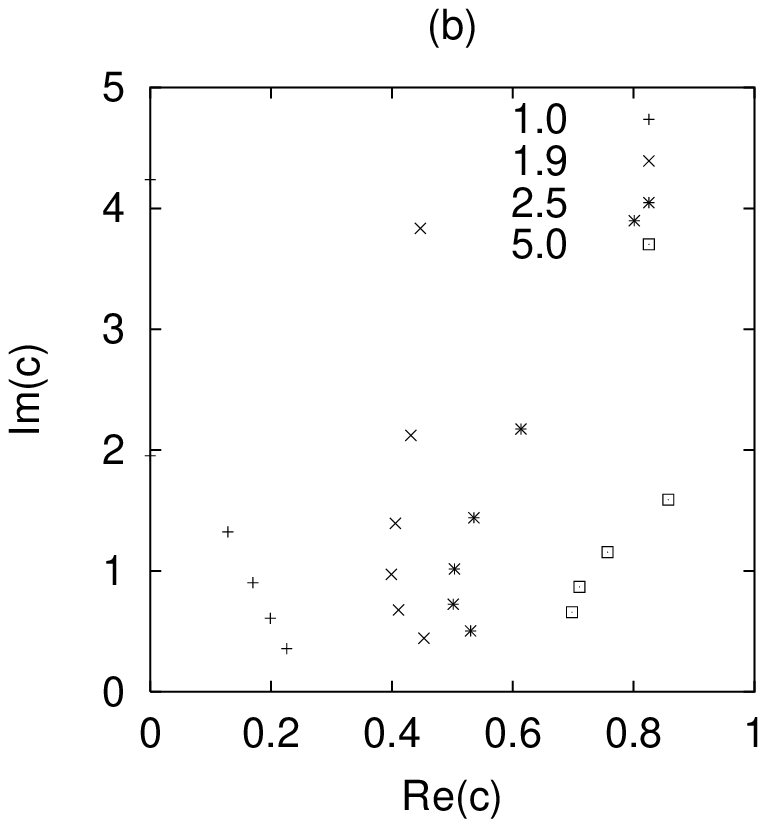}}
\centerline{\epsfxsize=3.0in \epsfysize=2.0in \epsfbox{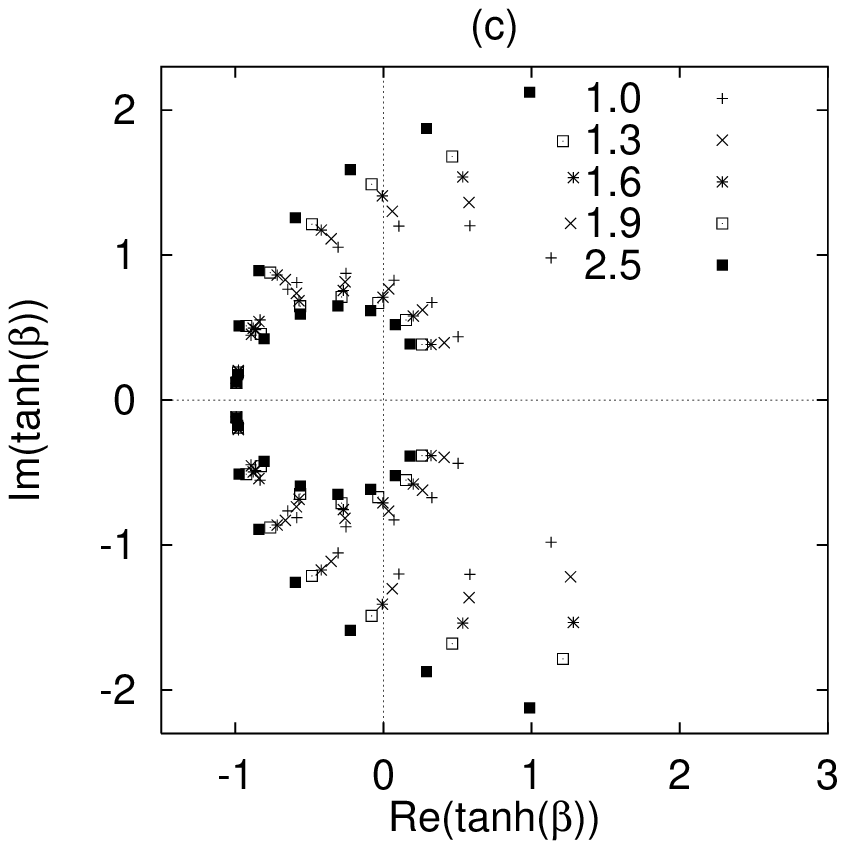}}
\caption{(a) Fisher zeros in the complex $c$ plane for a square dynamical 
lattice of size $N=14$ and at varying magnetic field 
($y=\e^{-2H}= 1.0,\ 1.6,\ 2.5$). For zero magnetic field, the zeros 
with ${\rm Re}(c)>0$ end in the vicinity of the physical critical point 
$c=(1/4,0)$.  
(b) The first quadrant in (a) shown magnified for several values of the 
fugacity $y=1.0,\ 1.6,\ 2.5,\ 5.0$.
(c) The trajectories in (a) shown in the complex ${\tilde c}$-plane  
for zero and nonzero magnetic field. The fugacity takes the values 
$y=1.0,\ 1.3,\ 1.6,\ 1.9,\ 2.5$ and $N=14$.} 
\label{f:4}
\end{figure}

We note from Fig.~\ref{f:3} and Fig.~\ref{f:4} that also the Fisher
zeros form curves in the complex plane.  In Fig.~\ref{f:3}(a) we
display the approach of the Fisher zeros for $H=0$ to the (physical
and unphysical) critical points on the real axis, $c=\pm 1/4$ for
increasing lattice size. There are also Fisher zeros on the imaginary
axis, which flow to $c=\pm i \infty$. In Fig.~\ref{f:3}(b) we show the
same curves mapped onto the $\tilde c$-plane, where the tilde refers
to the {\it dual} spin model. The usual duality relation is given by
$\tilde c ={\rm tanh}(\beta)$. We note that our model is not
self--dual, so the critical point $c_c = 1/4$ is different from the
dual critical point $\tilde{c}_c = 3/5$. Fig.~\ref{f:3}(b) suggests
the existence of an antiferromagnetic phase transition for the Ising
model on a square lattice with spins placed on the vertices
\cite{BJ}. It happens at $\tilde{c}_c^{({\rm af})} = 5/3$. This corresponds to
the unphysical value of $c=-1/4$, reflecting the fact that no 
antiferromagnetic transition occurs for the Ising model coupled to $2d$ 
gravity on a square lattice with spins placed on the faces. We notice
that the zeros in the $\tilde{c}$--plane are mapped {\it exactly} onto
each other under the transformation $\tilde{c}\to
1/\tilde{c}$. This implies that the critical exponents of the
antiferromagnetic transition are identical to the ones of the
ferromagnetic transition, as was noted in \cite{BJ}.
 
Fig.~\ref{f:4}(a) shows the flow of Fisher zeros with varying magnetic
field $H$ for a fixed lattice size $N=14$. The zeros on the arcs flow
away from the imaginary axis as $H$ increases, while the zeros on the
imaginary axis again appear to move towards $c=\pm i \infty$ (note
that these points are mapped onto $\tilde{c}=(-1,0)$ in
Fig.~\ref{f:4}(c)). For $H=0$ the zeros should pinch the real axis
like in Fig.~\ref{f:3}, but for $H\neq 0$ they seem to avoid the real
axis. This is compatible with the absence of a phase transition in the
presence of a symmetry breaking field. In Fig.~\ref{f:4}(c) we show
the same flow in the $\tilde c$-plane.

\subsection{Multihistogramming of Monte Carlo data}
\label{ss:imult}

For systems with more than a few spins, the study of complex zeros
must rely on numerical methods. Monte Carlo methods are very powerful
in studying a system using the full hamiltonian. They can be used to
study singularities of observables in the complex plane as was done in
the pioneering work ref.~\cite{MCZ}.  With the development of (single)
histogramming techniques the partition function could be calculated
for a continuous region in the coupling space and its zeros determined
in the complex plane \cite{MAR}. Using multihistogramming \cite{SW},
where Monte Carlo data taken at different couplings are combined
optimally, a more accurate determination of the partition function is
possible over a wide range of couplings. This method for computing
complex zeros was first used in \cite{ABV}. Here we provide a brief
description of the method, in order to describe the procedure we
followed. For details see \cite{HMP}.

The Monte Carlo updating of the triangulations was performed by the
so{--}called flip algorithm and the spins were updated using a
standard cluster algorithm. One updating sweep of the lattice
consisted of approximately $N_l$ {\it accepted} flips where $N_l$ is
the number of links of the triangulated surface.  After a sweep of the
lattice we updated the spin system.  All this is by now standard and
we refer to \cite{bj,atw} for details. We use the high quality random
number generator RANLUX \cite{mlfj} whose excellent statistical
properties are due to its close relation to the Kolmogorov K{--}system
originally proposed by Savvidy {\it et al.} \cite{ssa}.

The lattice sizes that we simulated ranged from 32 to 256 vertices.
In order to minimize finite size effects, we took the spins to be on
the vertices of the triangulations and included degenerate
triangulations with double links and vertices of order one \cite{atw}.
First we made a rough map of the partition function. Since the
partition function is invariant under $\beta\to\beta+i k \pi/2$ and
$H\to H+i k \pi/2$ it is not necessary to calculate $|Z_N|^2$ for
${\rm Im}(\beta) > \pi/2$ or ${\rm Im}(H)>\pi/2$.  In our calculation
we took $H$ to be purely imaginary.  Then we scanned the region where
we expected to see the first zeros, and ultimately we took a denser
grid of measurements around the points where the partition function
touched zero within errors. For the largest lattice, 55 million sweeps
were performed at each of 11 values $\beta_k$ around the location of
the zeros, in order to get a sufficiently accurate determination of
the partition function in this region. For each $\beta_k$ we computed
energy and magnetization histograms $h^k(E,M)$, where $E$ is the total
energy and $M$ the (absolute value of the) magnetization, from which
one can obtain an approximate estimate of the density of states:
\beq{*19}
\rho(E,M) = \sum_{k} w_k(E) \rho_k(E,M) = 
\sum_{k} w_k(E) h^k(E,M) \e^{\beta_k E}\, .
\eeq
The coefficients $w_k(E)$ are the appropriate weights for an optimal
determination of $\rho(E,M)$ 
\beq{*20}
w_k(E,M) = \frac{1}{
\sum_{l} \frac{n^l}{n^k}\frac{\tau_k}{\tau_l}
\exp\left((\beta_k-\beta_l) E + F_l\right) } \, ,
\eeq
where $n^{k,l}$ is the number of measurements at $\beta_{k,l}$, and
$\tau_{k,l}$ the respective autocorrelation times. $F_l$ is the free
energy at $\beta_l$, determined self--consistently from
\beq{*21}
\e^{-F_l}\equiv Z(\beta_l) =
\sum_{E,M,k} \frac{h^k(E,M)}{
\sum_j \frac{n^j}{n^k}\frac{\tau_k}{\tau_j}
\exp\left( (\beta_l-\beta_j) E + F_j\right) }\, .
\eeq
Then the partition function $Z(\beta)$ and any observable $\vev{{\cal
O}(E,M)}_\beta$  for real $\beta$ is  given by:
\bea{*22}
Z(\beta) &=& \sum_{E,M} \rho(E,M) \e^{-\beta E}\\
\vev{{\cal O}(E,M)}_\beta &=& 
\frac{1}{Z(\beta)}\sum_{E,M}{{\cal O}(E,M) \rho(E,M) \e^{-\beta E}}\, .
\eea
Lee--Yang zeros are computed from the minima of
\beq{*23}
\frac{Z(\beta,H)}{Z(\beta)} = 
\vev{\cos({\rm Im}(H)\,M)} - i \vev{\sin({\rm Im}(H)\,M)}
\eeq
for real $\beta$ and imaginary $H$ and $Z(\beta)\equiv
Z(\beta,0)$. The Fisher zeros were computed from the minima of
\beq{*24}
\frac{Z(\beta)}{Z({\rm Re}\beta)} = 
\vev{\cos({\rm Im}(\beta)\,E)} - i \vev{\sin({\rm Im}(\beta)\,E)}\, 
\eeq
for complex values of $\beta$ and $H=0$.
The errors in the partition function and in the position of the
complex zeros were computed by a standard binning procedure.

Table 1 and Table 2 contain our results for the zeros. We first
checked that multihistogramming was working quite well with our data
by looking at multihistograms of the specific heat and
susceptibility. It was relatively easy to observe the first few
Lee--Yang zeros. For Fisher zeros the computation was harder and in a
similar way we could only clearly observe the zero closest to the real
$\beta$ axis. Delicate cancellations in $|Z|^2$ between contributions
from the two terms in eq. \rf{*24} make the zeros located further away
from the real axis disappear in the statistical noise. This sets
limitations to the size of the surface on which we were able to
observe Fisher zeros.

$$\vbox {\offinterlineskip
\halign  { \strut#& \vrule# \tabskip=.5cm plus1cm
& \hfil#\hfil
& \vrule# & \hfil# \hfil &
& \vrule# & \hfil# \hfil &\vrule# \tabskip=0pt \cr \noalign {\hrule}
&& $N_v$ && $\beta\pm \Delta\beta$ && ${\rm Im}(H) \pm \Delta{\rm Im}(H)$ & \cr \noalign {\hrule}
&& 64 &&  0.1718 $\pm$ 0.0005  &&    0.203 $\pm$ 0.003 & \cr \noalign {\hrule}
&& 96 &&  0.1666 $\pm$ 0.0004  &&    0.1484 $\pm$ 0.0002 & \cr 
&&    &&  0.2114 $\pm$ 0.0004  &&    0.2454 $\pm$ 0.0004 & \cr 
\noalign {\hrule}
&& 128 &&  0.1643 $\pm$ 0.0014  &&    0.1182 $\pm$ 0.0005 & \cr
&&     &&  0.2037 $\pm$ 0.0012  &&    0.195 $\pm$ 0.005 & \cr 
&&     &&  0.2386 $\pm$ 0.0018  &&    0.260 $\pm$ 0.003 & \cr \noalign {\hrule}
&& 256 &&  0.1596 $\pm$ 0.0010  &&    0.069 $\pm$ 0.003 & \cr 
&&     &&  0.1850 $\pm$ 0.0017  &&    0.112 $\pm$ 0.004 & \cr 
&&     &&  0.2069 $\pm$ 0.0009  &&    0.148 $\pm$ 0.005 & \cr 
&&     &&  0.2262 $\pm$ 0.0009  &&    0.185 $\pm$ 0.001 & \cr 
&&     &&  0.2497 $\pm$ 0.0034  &&    0.175 $\pm$ 0.005 & \cr \noalign {\hrule}
}}$$ \nopagebreak
\vskip-0.5cm \nopagebreak
\begin{center} {\bf Table~1.} {\it Lee--Yang zeros observed using 
multihistogramming.}
\end{center}

$$\vbox {\offinterlineskip
\halign  { \strut#& \vrule# \tabskip=.5cm plus1cm
& \hfil#\hfil
& \vrule# & \hfil# \hfil &
& \vrule# & \hfil# \hfil &\vrule# \tabskip=0pt \cr \noalign {\hrule}
&& $N_v$ && ${\rm Re}(\beta)\pm \Delta{\rm Re}(\beta)$ && ${\rm Im}(\beta) \pm \Delta{\rm Im}(\beta)$ & \cr \noalign {\hrule}
&& 32 &&  0.1428 $\pm$ 0.0016  &&  0.1773 $\pm$ 0.0005 & \cr \noalign {\hrule}
&& 64 &&  0.1481 $\pm$ 0.0049  &&  0.1348 $\pm$ 0.0022 &  \cr \noalign {\hrule}
&& 96 &&  0.1533 $\pm$ 0.0021  &&  0.1121 $\pm$ 0.0034 &  \cr \noalign {\hrule}
&& 128 && 0.1556 $\pm$ 0.0300  &&  0.1030 $\pm$ 0.0050 &  \cr \noalign {\hrule}
&& 256 && 0.1527 $\pm$ 0.0035  &&  0.0788 $\pm$ 0.0047 &  \cr \noalign {\hrule}
}}$$ \nopagebreak
\vskip-0.5cm \nopagebreak
\begin{center} {\bf Table~2.} {\it Fisher zeros observed using 
multihistogramming.}
\end{center}

\section{Verification of scaling relations}
\label{s:v}

In this section we discuss the extent to which the scaling relations
eqs.~(\ref{*8}--\ref{*14}) hold for the Ising model on a {\it dynamical}
lattice. There is no {\it a priori} reason to expect these relations
to hold. As discussed earlier, for the fixed lattice, their validity is due to
the divergence of the spin--spin correlation length in the
critical region. For the Ising model on a dynamical lattice, we
expect the correlation length associated to {\it geometry} fluctuations to
diverge \cite{aw,syracuse,ajw}, but the same need not necessarily be true
for the spin--spin correlation length even at a continuous transition
\cite{ha}. Recently however, numerical evidence indicated 
that this is indeed the case, and that the system behaves as an
ordinary statistical system near a third order phase transition
\cite{corr}.

Numerical simulations \cite{syracuse,ajw,corr} further indicate that
the Hausdorff dimension of the system is very close to $4$.  From this
value we can estimate the linear size $L\equiv N^{1/d_H}$ of the
systems that we are studying and find that it is indeed quite
small. In spite of substantial finite size effects, our results will
provide evidence that the scaling relations are indeed satisfied by
the motion of the complex zeros with varying lattice size, couplings,
or the order of the zero. We can, in general, observe reasonable
agreement with the known critical exponents (we will refer to these as
the KPZ exponents).  In the cases where a deviation from the expected
value is observed, we can observe an asymptotic approach to this value
with increasing lattice size.

\subsection{Zeros of the exact partition functions for small
square lattices}

\begin{figure}[htb]
\centerline{\epsfxsize=4.0in \epsfysize=2.67in \epsfbox{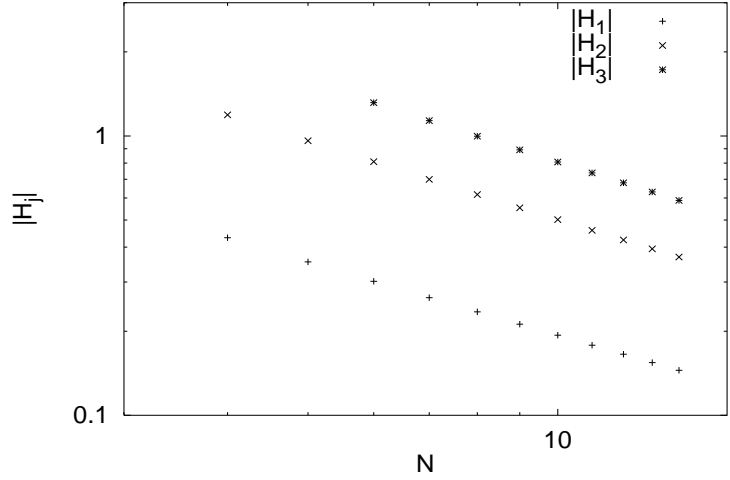}}
\caption{Scaling of the first three Lee--Yang zeros for the Ising model on
square dynamical lattices of size $N$ ($j$ labels the zero). The slopes are
expected to be given by the exponent combination 
$-\beta\delta/(\nu d_H)=-5/6$ ($d_H$ is the Hausdorff dimension).
The fitted slopes are given in Table~3.}
\label{f:5}
\end{figure}

At $K=0$ and large enough $N$ eq.~\rf{*9} implies
\begin{equation}
\ln |H_j| = -{{\beta\delta}\over{\nu d_H}} \ln(N) + C_j \, .\label{eq:216}
\end{equation}
The slope on a log--log plot of $H_j$ vs. $N$ is expected to take the value 
\begin{equation}
-{{\beta\delta}\over{\nu d_H}} = -5/6 = -0.8333... 
\label{eq:errcom}
\end{equation}
for all $j$. Fig.~\ref{f:5} and Table~3 show our results for the first
three Lee--Yang zeros. The extracted exponent combination is in
reasonable agreement with the KPZ exponents, especially for the first
zero.  The errors reported in Table~2 are not true statistical errors
(which are meaningless in this case). They are computed from the
standard formula giving least squares linear fit errors and in this
case they are simply a measure of the systematic deviation of the
points from a straight line. We will follow this practice for our
fits throughout this section.

$$\vbox {\offinterlineskip
\halign  { \strut#& \vrule# \tabskip=.5cm plus1cm
& \hfil#\hfil
& \vrule# & \hfil# \hfil &
&\vrule# \tabskip=0pt \cr \noalign {\hrule}
&& $j$ && Slope  & \cr \noalign {\hrule}
&& 1 && -0.871 $\pm$ 0.002 & \cr \noalign {\hrule}
&& 2 && -0.935 $\pm$ 0.002 & \cr \noalign {\hrule}
&& 3 && -0.951 $\pm$ 0.002 & \cr \noalign {\hrule}
}}$$ \nopagebreak
\vskip-0.5cm \nopagebreak
\begin{center} {\bf Table~3.} 
{\it Predictions for the combination of critical exponents 
$-{{\beta\delta}\over{\nu d_H}}$ from scaling of \\
the first three Lee--Yang zeros of exactly known partition functions $Z_N$.}
\end{center}

\begin{figure}[htb]
\centerline{\epsfxsize=4.0in \epsfysize=2.67in \epsfbox{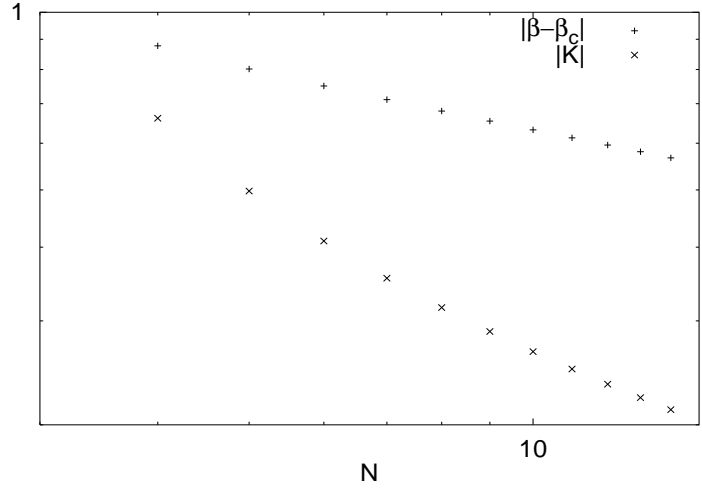}}
\caption{Illustration of the scaling relation $|K|\propto N^{-1/(\nu d_H)}$ for
the first Fisher zero at $H=0$. Due to finite size effects, the curve for $|K|$
is not a straight line. On the other hand, $|\beta -\beta_c|$ scales
well and yields a slope of $-0.327 \pm 0.01$. The expected slope is $-1/3$.}
\label{f:6}
\end{figure}

Eq.~\rf{*14}  implies
\begin{equation}
\ln|K_j| \equiv \ln|u_j-u_c| = {1\over{\nu d_H}}\ln \left( \frac{j}{N}\right) 
+ C\ .
\label{eq:217}
\end{equation}
Similarly, since 
$|K_j|\propto |\beta_j-\beta_c| + O(|\beta_j-\beta_c|^2)$, we would expect that
\begin{equation}
\ln|\beta_j-\beta_c| = {1\over{\nu d_H}}\ln \left( \frac{j}{N}\right) + C\ , 
\label{eq:be}
\end{equation}
where the slope should be given by 
\begin{equation}
{1\over{\nu d_H}} = {1\over{3}}\, ,\ \label{eq:xxx}
\end{equation}
and the constant $C$ should be independent of $j$.  Table~4 shows the
results of fits to eq.~(\ref{eq:be}), and similar fits with $N$
replaced by $N_v$, the number of vertices of the lattice.  The
corresponding fits using the scaling variable $K$ did not yield
straight lines in a log--log plot, as can be seen from Fig.~\ref{f:6}.
In this figure, we have plotted $|K_1|$ (lower curve)
resp. $|\beta_1-\beta_c|$ (upper curve) vs. $N$ on a logarithmic
scale. The difference between the scaling behaviour of the two
different scaling variables (which are expected to be identical for
very large systems), shows that we are not deep in the scaling region
and that finite size effects are important for the system sizes that
we consider. The same must be said about the difference between the
slopes and intercepts $C$ of Table~4. From eq.~\rf{eq:be} it is
expected that the slopes and $C$ should take the same value for all
$j$, so that the data points would fall on one and the same curve. The
difference between the results obtained by using $N$ and $N_v$ should
asymptotically decrease as the lattice size goes to infinity. (In the
table, $n$ denotes the number of points included in the fit.)

$$\vbox {\offinterlineskip
\halign  { \strut#& \vrule# \tabskip=.5cm plus1cm
& \hfil#\hfil
& \vrule# & \hfil# \hfil &
\vrule# & \hfil# \hfil &
\vrule# & \hfil# \hfil &
\vrule# & \hfil# \hfil &\vrule# \tabskip=0pt \cr \noalign {\hrule}
&& Quantity fitted && $j$ && Slope && $C$ && $n$ & \cr \noalign {\hrule}
&& $\ln|\beta-\beta_c|$ vs. $\ln(j/N)$   && 1 && 0.327 $\pm$ 0.001 && 
0.295 $\pm$ 0.002 && 8 & \cr 
&&                                       && 2 && 0.377 $\pm$ 0.002 && 
0.490 $\pm$ 0.005 && 4 & \cr 
\noalign {\hrule}
&& $\ln|\beta-\beta_c|$ vs. $\ln(j/N_v)$ && 1 && 0.383 $\pm$ 0.002 && 
0.492 $\pm$ 0.006 && 6 & \cr 
&&                                       && 2 && 0.438 $\pm$ 0.004 && 
0.667 $\pm$ 0.009 && 4 & \cr 
\noalign {\hrule}
}}$$ \nopagebreak
\vskip-0.5cm \nopagebreak
\begin{center} 
{\bf Table~4.} {\it Slopes and intercepts from the fits to eq.~(\ref{eq:be}). 
The theoretical value for the \\ slope is  ${1\over{\nu d_H}}=1/3$. 
The dependence of the slope and of $C$ on $j$ indicates finite size effects,\\
as does the difference between $N$ and $N_v$. $n$ is the number of
degrees of freedom in the fits.}
\end{center}

\begin{figure}[htb]
\centerline{\epsfxsize=3.0in \epsfysize=2.0in \epsfbox{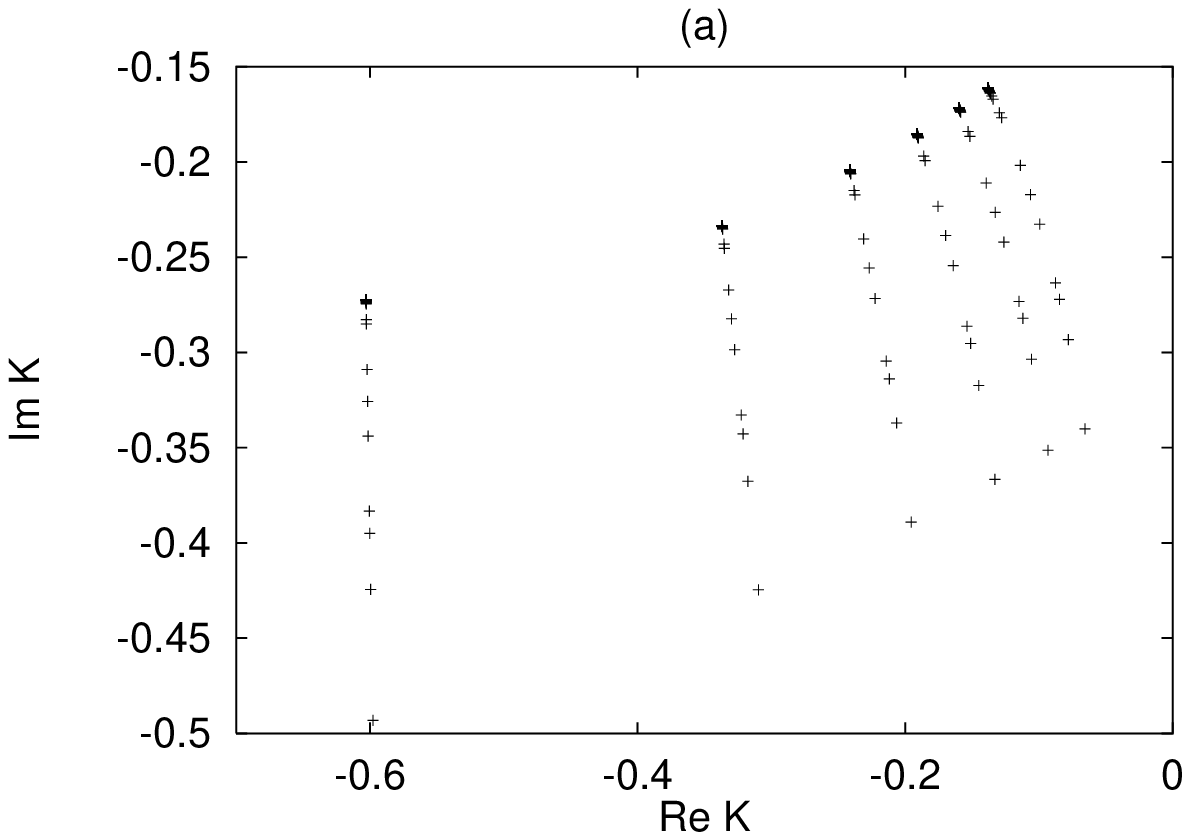}}
\centerline{\epsfxsize=3.0in \epsfysize=2.0in \epsfbox{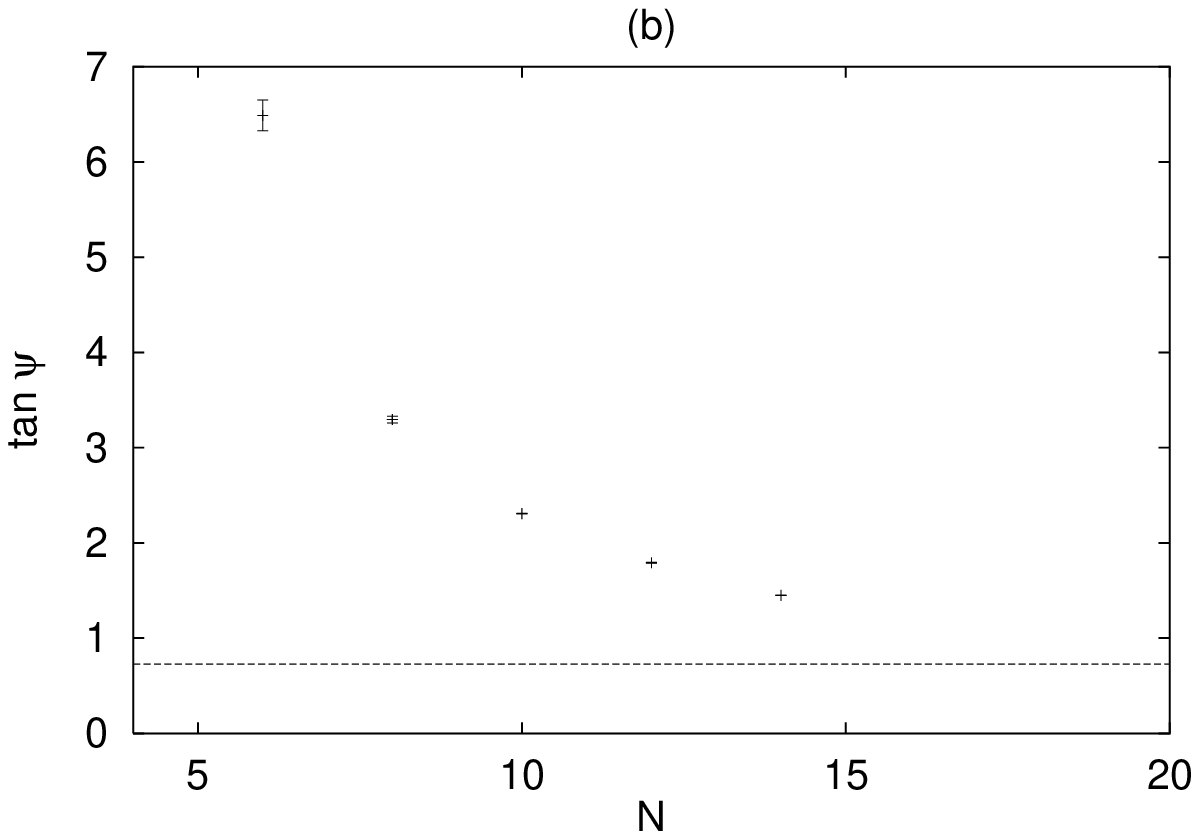}}
\centerline{\epsfxsize=3.0in \epsfysize=2.0in \epsfbox{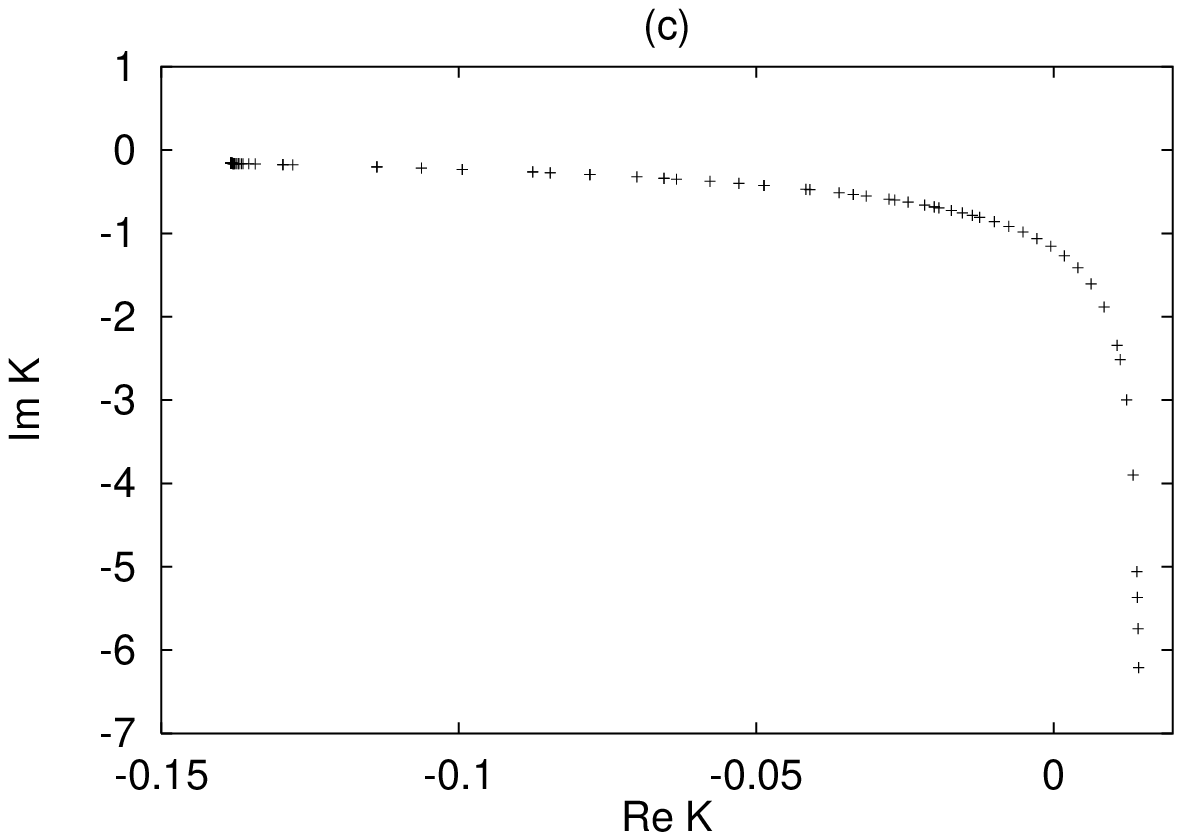}}
\caption{(a) Trajectories of the first Fisher zero in the complex $K$ 
plane for nonzero magnetic field $-0.2<{\rm Im}(H)<0.2$. 
The trajectories are for small square lattices of size 
$N=4,\ 6,\ 8,\ 10,\ 12,\ 14$ in order from left to right. The angle
$\psi$ that the trajectories form with the ${\rm Re}(K)$ axis is
expected to approach $\psi = \pi / (2 \beta \delta) = 36^{\circ}$ as
$N\to \infty$.  (b) ${\rm tan}\psi$ versus $N$ for small
lattices. The dashed line corresponds to the expected asymptotic value
of ${\rm tan}\psi$ for large $N$.  (c) Breakdown of scaling for
strong magnetic field (up to $|H|\approx 2.6$). The angle $\psi$ changes
drastically for strong $H$.}
\label{f:7}
\end{figure}

The scaling relation eq.~\rf{*11}
predicts the angle that the trajectories of the zeros 
will form with respect to the ${\rm Re}(K)$ axis as $H$ is varied for
large $L$. In our case the angle is expected to be
\begin{equation}
\label{*25}
\psi = {{\pi}\over{2\beta\delta}} = 36^{\circ}\ .
\end{equation}
This prediction is valid for large system sizes $L$ and small magnetic field.
In Fig.~\ref{f:7}(a) we show the trajectories of the first Fisher zero for
$N=4,\ 6,\ 8,\ ...,\ 14$ and magnetic fields in the interval
$-0.2 < H < 0.2$. As expected from eq.~\rf{*11}, $H=0$ corresponds to 
the ``turning point'' closest to the ${\rm Re}(K)$ axis.
Fig.~\ref{f:7}(b) and Table~5 show the corresponding values of
$\tan\psi$, calculated from fits in the small $H$ range $-0.03<H<0.02$. 
Although $\psi$ does not reach its infinite volume value,
Fig.~\ref{f:7}(b) shows the approach to it with
increasing lattice size. The breakdown of 
scaling for very strong $H$ can be observed in Fig.~\ref{f:7}(c) where a
trajectory for $|H|$ between zero and 2.6 are shown. 

$$\vbox {\offinterlineskip
\halign  { \strut#& \vrule# \tabskip=.5cm plus1cm
& \hfil#\hfil
& \vrule# & \hfil# \hfil &
&\vrule# \tabskip=0pt \cr \noalign {\hrule}
&& $N$ && ${\rm tan}\psi\pm\Delta{\rm tan}\psi$  & \cr \noalign {\hrule}
&& 6  && 6.489 $\pm$  0.162 & \cr \noalign {\hrule}
&& 8  && 3.295 $\pm$  0.036 & \cr \noalign {\hrule}
&& 10 && 2.308 $\pm$  0.001 & \cr \noalign {\hrule}
&& 12 && 1.792 $\pm$  0.006 & \cr \noalign {\hrule}
&& 14 && 1.450 $\pm$  0.002 & \cr \noalign {\hrule}
&& $\infty$ && 0.727          & \cr \noalign {\hrule}
}}$$ \nopagebreak
\vskip-0.5cm \nopagebreak
\begin{center} {\bf Table~5.} 
{\it Measured $(N=6,\ 8,\ ...,\ 14)$ and expected (eq.~\rf{*25}) slopes of \\ 
trajectories of zeros in the complex $K$ plane as the magnetic field $H$  
is varied, \\ as a function of the size of the lattice.}
\end{center}

\begin{figure}[htb]
\centerline{\epsfxsize=4.0in \epsfysize=2.67in \epsfbox{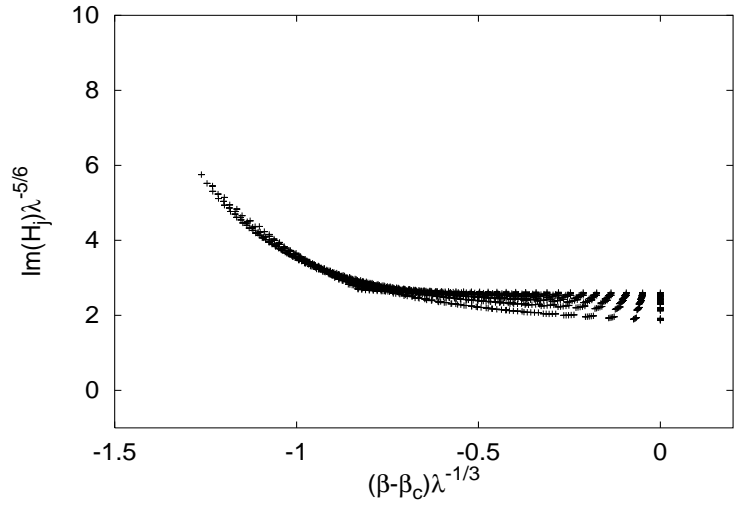}}
\caption{Check of the scaling relation for the $j$th Lee--Yang zero
$H_j^2\lambda^{-2\delta/(\delta + 1)} = F(K \lambda^{-1/(\nu d_H)})$,
where $\lambda=j/N$ and $F$ is a universal function. All values of
$j>1$ for lattice sizes in the range $10 \le N\le 14$ have been
plotted on the same graph. The points are expected to fall on the same
universal curve for {\it large $j$}.}
\label{f:9}
\end{figure}

We also checked the validity of the scaling relation given by
eq.~\rf{*13}. Taking $K$ real, we plotted $H_j(N/j)^{5/6}$
vs. $(\beta-\beta_c)(N/j)^{1/3}$.  The scaling function $F$ is shown
in Fig.~\ref{f:9}.  All zeros of order $j>1$ for lattice sizes ($10\le
N\le 14$) are included in the same graph.  We observe that there
exists a range of $\beta$ in the hot phase where the data points lie
approximately on a univeral curve defining the function $F$.

By varying the value of the exponent $-\delta/(\delta + 1)=-5/6$ of
the scaling variable $\lambda = j/N$ on the $y$-axis away from its
KPZ value, we observed a broadening of the curve and in this way we
obtained a (somewhat subjective) determination of the exponent
combinations
\begin{equation}
{\delta\over{\delta+1}} = {{\beta\delta}\over{\nu d_H}} \approx 0.85
\pm 0.05 \, , 
\end{equation}
(the equality of these follow from the general exponent equalities and was 
used in the derivation of the scaling relation \rf{*13}).  
This is in excellent agreement with the KPZ value of $5/6$.

\subsection{Zeros determined from multihistogramming}

In this section we describe the scaling of the zeros determined
numerically from Monte Carlo data. The lattice sizes investigated here
range from 32 to 256 vertices (60--508 triangles).  We were able to
determine the first five Lee--Yang zeros for the biggest lattice size,
while for the smaller ones at most three zeros were visible.  
For the Fisher zeros, only the zero closest to the real axis 
was observed for any lattice size.
       
The Lee--Yang zeros we observed with multihistogramming are listed in
Table~1.  The least-squares fits to eq.~\rf{eq:216} give for the
first and second Lee--Yang zero,
\begin{eqnarray}
{{\beta\delta}\over{\nu d_H}} &=& 0.773 \pm 0.013 \ \ (j=1)\\
{{\beta\delta}\over{\nu d_H}} &=& 0.788 \pm 0.033 \ \ (j=2)\, .
\label{eq:LeeYY}
\end{eqnarray}
All data points with $j\le 2$ shown in Table~1 were included in the fits. 
Using $N_v$ instead of $N$ for the volume of the system we obtain:
\begin{eqnarray}
{{\beta\delta}\over{\nu d_H}} &=& 0.787 \pm 0.013 \ \ (j=1)\\
{{\beta\delta}\over{\nu d_H}} &=& 0.800 \pm 0.034 \ \ (j=2)\, ,
\label{eq:Lee-Y}
\end{eqnarray}
which gives a measure of the presence of finite size effects.  The
results are in quite good agreement with the expected value of
${\beta\delta}/{\nu d_H}=5/6$. From the exactly known value $\nu d_H =
3$ we obtain
\begin{eqnarray}
\beta\delta &=& 2.36 \pm 0.04 \ \ \ (j=1) \\
\beta\delta &=& 2.40 \pm 0.10 \ \ \ (j=2) \ .
\end{eqnarray}
The exact value of this combination of exponents is 2.5.

\begin{figure}[b]
\centerline{\epsfxsize=4.0in \epsfysize=2.67in \epsfbox{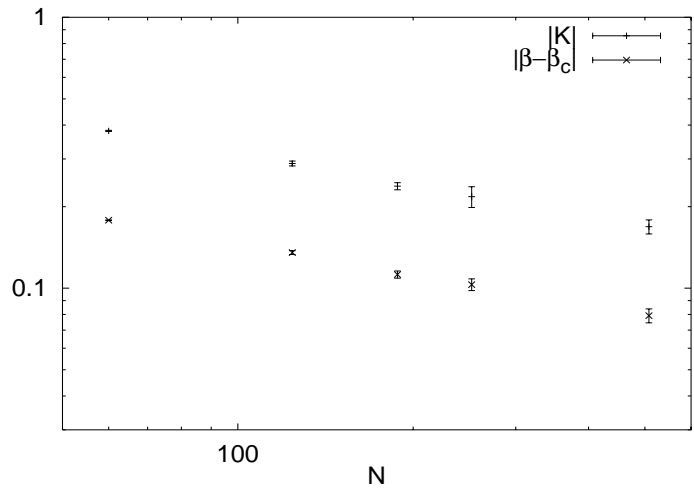}}
\caption{The first Fisher zero observed using
multihistogramming on dynamical triangular lattices in the size range
$32\le N_v \le 256$ ($60\le N \le 508$). The slopes of the lines are
expected to be $-1/(\nu d_H)=-1/3$. }
\label{f:11}
\end{figure}

The value of $\nu d_H$ extracted from eq.~\rf{eq:217} is somewhat
lower than 3, but still remarkably close to it given the small size of
the systems considered. The plot of the observed zeros is shown in
Fig.~\ref{f:11} ($\ln|K|$ resp.  $\ln|\beta-\beta_c|$
vs. $\ln(N)$). We also calculated the slope of $\ln|K|$ vs. $\ln(N_v)$
and of $\ln|\beta-\beta_c|$ vs. $\ln(N_v)$.  The results are shown in
Table~6. We note that by discarding the smallest lattices the value of
$\nu d_H$ approaches 3, albeit with an increasing error. For example,
including only the three largest lattices we obtain (using $N$ for
volume):
\beq{*flg}
\frac{1}{\nu d_H} = 0.350 \pm 0.067 \, .
\eeq

$$\vbox {\offinterlineskip
\halign  { \strut#& \vrule# \tabskip=.5cm plus1cm
& \hfil#\hfil
& \vrule# & \hfil# \hfil &
&\vrule# \tabskip=0pt \cr \noalign {\hrule}
&& Quantity fitted && $1/(\nu d_H)$ & \cr \noalign {\hrule}
&&  $\ln|K|$ vs. $\ln(N)$  && 0.392 $\pm$ 0.016 & \cr \noalign {\hrule} 
&&  $\ln|K|$ vs. $\ln(N_v)$    && 0.407 $\pm$ 0.017 & \cr\noalign {\hrule} 
&& $\ln|\beta-\beta_c|$ vs. $\ln(N)$ && 0.386 $\pm$ 0.014 & \cr \noalign 
{\hrule}
&& $\ln|\beta-\beta_c|$ vs. $\ln(N_v)$   && 0.401 $\pm$ 0.014 & \cr \noalign 
{\hrule}
}}$$ \nopagebreak
\vskip-0.5cm \nopagebreak
\begin{center} 
{\bf Table~6.} {\it Finite size scaling for the first Fisher zero.}
\end{center}

\section{Conclusion} 
\label{s:c}

We have computed the positions of the singularities of the partition
function for the two--dimensional Ising model coupled to gravity in
the complex plane. Both Lee--Yang and Fisher zeros were studied.

We verified the result in \cite{Staudacher} that the Lee--Yang zeros
are located on the unit circle of the complex fugacity plane. This
presents us with the challenge of proving a corresponding Lee--Yang
theorem for the case when a fluctuating metric contributes an
additional quantum degree of freedom. We also observed that the Fisher
zeros form one dimensional curves in the complex temperature plane.

Given the small size of the systems we studied, we obtained reasonable
agreement with scaling laws derived from ordinary renormalization
group arguments. Although one cannot use our results to accurately
determine the scaling exponents, the extracted exponent combinations
show a reasonable agreement with their exactly known values
\cite{BK}. This, together with the fact that a diverging matter
correlation length exists in the system \cite{corr}, gives us
confidence to conclude that the critical behaviour of the system is
given by the scaling of the distribution of the complex zeros of the
partition function.

\newpage


\clearpage

\begin{figure}[b]
\centerline{\epsfxsize=4.0in \epsfysize=2.67in \epsfbox{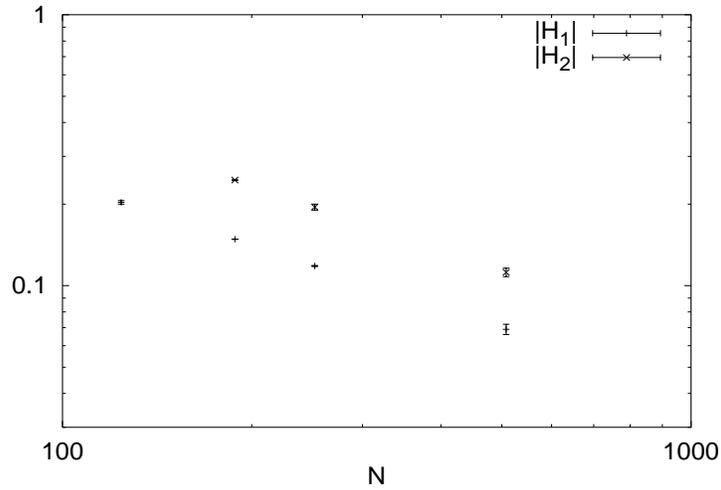}}
\caption{The first ($j=1$) and second ($j=2$) Lee--Yang zero observed 
using multihistogramming on dynamical triangular lattices in the size range
$64\le N_v\le 256$ ($124\le N \le 508)$. The slopes of the lines
should be given by $-\beta\delta / (\nu d_H) = -5/6$.}
\label{f:10}
\end{figure}

\end{document}